**Generative Artificial Intelligence Adoption Among Bangladeshi Journalists: Exploring Journalists' Awareness, Acceptance, Usage, and Organizational Stance for GenAI**

**Author Note**
H M Murtuza (corresponding author)
ORCID: https://orcid.org/0009-0009-6774-7604
Email: hmmurtuza@ou.edu, LinkedIn
Md Oliullah
ORCID: https://orcid.org/0009-0003-7493-4762
Email: md.oliullah-1@ou.edu, LinkedIn
H M Murtuza and Md Oliullah are Ph.D. Students at Gaylord College of Journalism and Mass Communication, University of Oklahoma.

**Abstract**

Newsrooms and journalists across the world are adopting Generative AI (GenAI, but commonly referred to as AI). Drawing on in-depth interviews of 23 journalists, this study identifies Bangladeshi journalists' awareness, acceptance, usage patterns, and their media organizations' institutional stance toward GenAI. This study finds Bangladeshi journalists' high reliance on GenAI as like their western colleagues despite limited institutional support and almost complete absence of AI policy. Despite this stark contrast, the concerns over GenAI's implications in journalism between the West and non-West were mostly identical. Moreover, this study also contributes to the Unified Theory of Acceptance and Use of Technology (UTAUT) by proposing two major changes regarding the adoption of GenAI among journalists in non-Western settings. First, this study identifies the non-contribution of facilitating conditions in shaping behavioral intent in adoption of GenAI in non-Western contexts. Second, social influence works in a horizontal order through informal peer pressure or professional motivation in absence of vertical formal institutional hierarchical or peer pressure. Voluntariness in the context of Bangladeshi journalists is underpinned by their professional compulsion. Therefore, this study contributes to understanding how contextual factors shape technology adoption trajectories in non-Western journalism.

**Generative Artificial Intelligence Adoption Among Bangladeshi Journalists**

The emergence of generative artificial intelligence (GenAI), the latest form of AI, in recent years has been transformative for journalism, especially in the ways it engages with advanced digital technologies (Gondwe, 2024; Lewis et al., 2025; Ogbebor, & Carter, 2021; Onebunne, 2025; Simon, 2024; Thomas & Thomson, 2023). More specifically, the launch of ChatGPT, a breakthrough in GenAI, introduced a technology capable of generating content—text, images, audio, and video—based on natural language processing (Brown et al., 2020). The first widely known demonstration of the technology appeared in the form of an opinion article published by the British newspaper *The Guardian* on September 8, 2020, written by one of the earliest models, GPT-3, showcasing a significant advancement in the field of GenAI and its application in journalism (GPT-3, 2020).

Though in many ways AI or algorithmic tools were earlier involved in journalism, they were not used in any effective form of creative activities like writing op-eds. GenAI (e.g., ChatGPT or DeepSeek) has so far been employed in many ways, including the enhancement of quality, efficiency, and productivity among journalists (Albizu-Rivas et al., 2024; Adjin-Tettey et al., 2024; Cools & Diakopoulos, 2024; Verma, 2024). AI enabled journalists "to engage more deeply with complex stories, leveraging their human insights, empathy, and investigative skills in ways that were not possible before" (Verma, 2024).

Meanwhile, studies have taken place to understand the impacts of AI adoption in journalism—AI's efficiency in changing audience perception (Horne et al., 2019), AI's efficiency in news processing (Owsley & Greenwood, 2022), and its implications in journalism in general (Noain-Sánchez, 2022; Sun et al., 2022). Credibility and authorship, generation of manipulative content, misinformation, and privacy, among others, emerge as major concerns because of the widespread use of GenAI in journalism. Maintaining the autonomy of the media is another crucial challenge as the use of GenAI becomes widespread (Simon, 2024). However, understanding the role AI plays in newsrooms, especially in the news production phase—ideation, pre-reporting analysis, reporting, copy editing—is understudied, let alone studies on the non-Western, developing, and underdeveloped world (Simon, 2022 & 2024).

Bangladesh, representing a non-Western country from the Global South and a developing country, makes it a suitable site for this study. The country's sharp socio-economic and political disparity with the West further rationalizes the appropriateness of its selection for this study. This study covers Bangladesh's perspective from a broad albeit basic level, starting with the level of awareness, usage, potentials, implications, and institutional policy stance regarding GenAI in journalism. The importance of this study lies in its contribution to the greater discussions on GenAI adoption in journalism, especially in the news production phase, how the adoption in the non-Western world has been different from the West, and factors contributing to making the differences. Additionally, this study contributes to the Unified Theory of Acceptance and Use of Technology (UTAUT) (Venkatesh et al., 2003) by showing how adoption of GenAI can be different albeit quick among journalists in a non-Western context without institutional support, based on professional motivation when career threat is prominent. Identification of the non-contribution of some of the major elements of UTAUT in a non-Western context marks this study's contribution to scholarly discourse.

## Literature Review

### *Unified Theory of Acceptance and Use of Technology*

The Unified Theory of Acceptance and Use of Technology (UTAUT) framework proposed by Venkatesh et al. (2003) brings together eight information system models under a single umbrella to examine how new technology infusion and adoption progress within organizations. Prior to this, researchers relied on several dominant models from psychology, sociology, and IS management disciplines, which captured approximately 40% of the variance behind technology adoption. UTAUT was developed to better explain technology adoption by consolidating insights from these earlier models. The initial UTAUT was based on eight prominent theoretical models and was later expanded to include consumer behavior and cultural factors (Marikyan & Papagiannidis, 2025; Venkatesh et al., 2003; Venkatesh & Zhang, 2010). By integrating 32 separate constructs, the UTAUT framework increased explanatory power to approximately 70% of the variance in users' technology adoption decisions.

The first model of UTAUT was developed for mandatory and voluntary use of technology in an organizational setting (Venkatesh et al., 2003; Marikyan & Papagiannidis, 2025). Venkatesh et al. (2012) developed UTAUT2 to check technology adoption among the consumer segment of the population, particularly the private user segment and individuals in non-organizational settings, by incorporating three additional dimensions—hedonic motivation (fun or pleasure derived from using technology), price value (trade-off between the perceived benefit and the monetary cost), and habit (when people tend to perform behavior automatically). The voluntariness of technology use was dropped from the moderators like learning, age, gender, and experience. Grounding on UTAUT and UTAUT2, this qualitative study aims to identify the contributing factors behind journalists' adoption of AI in a South Asian context, specifically in the context of Bangladesh.

### Artificial Intelligence, Generative AI, and Journalism

While UTAUT provides a theoretical lens for understanding technology adoption, understanding the specific technologies and their applications in journalism is essential. The emergence of generative AI (GenAI) tools represents a particularly significant development for journalism practice. The term *Artificial Intelligence* was first proposed by John McCarthy and his team in 1955 (McCarthy et al., 1955). After them, many scholars have defined AI from varied perspectives (Boden, 2016; Broussard et al., 2019; Legg & Hutter, 2007; McCarthy, 2007; Nilsson, 1998). "The key to reaching human-level AI is making systems that operate successfully in the common sense informatic situation" (McCarthy, 2007, p. 1175). However, there is no unified definition of AI. From an operationalization standpoint, AI development refers to "activity devoted to making machines intelligent" (Nilsson, 1998, p. 1). Because of technological advancements, many forms of AI have emerged, and Generative AI is one of the subfields of AI. From a technical standpoint, GenAI leverages machine learning, neural networks, and other advanced techniques to produce new content (e.g., text, images, audio, video, and other forms of media) (Sengar et al., 2024; Ooi et al., 2025). GenAI also works based on large language models, learning the underlying patterns and structures through large data corpora and therefore allowing the model to generate fresh data of the same pattern (Sengar et al., 2024).

Since the launch of the LLM-based GenAI model ChatGPT, the technology has drawn huge interest among people. The reason behind the popularity of ChatGPT, DeepSeek, Google LM, and others is the accessibility and affordability of the technology due to their capability to process natural language-based commands (Helberger & Diakopoulos, 2023). Because affordability and the knowledge required for using AI are minimal, journalists in developing and non-Western countries like Bangladesh are heavily using AI technology.

According to Simon (2022), journalists adopt AI basically for several reasons: technological developments, market pressures, industry dynamics, and uncertainty, hype, and hope. The involvement of AI includes a wide range of tools and methods that are capable of carrying out different tasks of varying degrees of complexity, autonomy, and abstraction (Lewis & Simon, 2023). Based on different intentional forms of use and to distinguish between different AI capabilities, varied terms such as "algorithmic journalism" (Anderson, 2013), "automated journalism" (Nickolas, 2019), and "computational journalism" (Cohen et al., 2011) are assigned, representing the gradual adoption of AI in journalism by news outlets.

Journalists use AI for various tasks, including brainstorming, preparing interview questions, transcribing interview data, writing and drafting news, improving writing quality, and summarizing stories (Adjin-Tettey et al., 2024; Broussard et al., 2019; Chan-Olmsted, 2019; Jamil, 2021; Wu, 2024). While early adoption occurred in well-resourced organizations like *The New York Times*, *The Washington Post*, the *Associated Press*, and *Reuters*, the emergence of GenAI has democratized AI use by making it accessible and affordable for newsrooms of all sizes (Harb & Arafat, 2024; Sonni, 2025; Rayyan, 2025). This accessibility is particularly significant for journalists in developing countries like Bangladesh, where financial constraints have historically limited technology adoption.

### Challenges and Implication of AI use in Journalism

While AI tools offer significant benefits for journalism efficiency, the adoption of AI in newsrooms also presents substantial challenges and raises critical questions about journalism's fundamental values. Understanding both opportunities and risks is essential for responsible adoption. While AI can help journalists produce more personalized content and manage information overload, its adoption also raises ethical concerns about journalism practice. Biases in AI training data risk reinforcing existing biases, potentially compromising ethical reporting and eroding public trust in media (Newman, 2018; Wenger et al., 2024).

AI use can make journalism more convenient and cost-efficient, but it comes with some challenges, too. Though AI can help journalists and news organizations produce more personalized content and formats, along with managing information overload, the use of AI also raises existential and ethical questions (Newman, 2018). The biases AI might have in its training data might lead to the reinforcement of unintentional existing biases. These have the potential to hamper ethical and fair reporting and lead to a subsequent decline in public trust in media. Wenger et al. (2024) pointed out ethical concerns over the use of AI as a major burning issue stemming from the growing adoption of AI tools by journalists and media organizations. The impact of AI varies depending on the specific tasks being automated (Simon, 2022). While AI integration can improve efficiency and speed, it does not provide effectiveness in all contexts—for example, AI struggles with political nuance, making integration into news generation complex. Research shows audience concerns about AI-generated content: news readers perceived automated reports as lower quality than human-written reports (Zheng et al., 2018), and AI-generated headlines as less credible than human-written headlines (Longoni et al., 2022).

Organizational and professional challenges compound these concerns. Beckett (2019) found three major obstacles to AI adoption across 71 news organizations: financial resource constraints, lack of knowledge or skills, and cultural resistance including fears of job loss. Media professionals express particular distrust of AI tools due to limited knowledge, fear of redundancy, and concerns that AI will undermine journalism's core values (Noain Sánchez, 2022). Addressing these challenges requires education and training that builds critical thinking and overcomes misconceptions about AI (Noain Sánchez, 2022). As AI tools increasingly generate convincing content, audiences face growing challenges distinguishing authentic from false information, threatening public trust in news (Newman, 2023).

### AI Literacy

Given these challenges—limited knowledge, professional distrust, and quality concerns—AI literacy emerges as a critical factor in responsible adoption. Beyond simply operating AI tools, journalists need a comprehensive understanding of AI capabilities, limitations, and ethical implications. Literacy of new technology plays a vital role in understanding how informed people make decisions about using technologies. Like other technologies, AI or GenAI literacy also contributes to shaping individuals' terms of engagement with the tools as they become aware of their pros and cons based on knowledge. In 2016, Konishi Yoko first introduced AI literacy as a subset of digital literacy (Yoko, 2016; Mansoor et al., 2024). However, a comprehensive definition of AI literacy is still in the process of evolving. Steinbauer et al. (2021)

treat AI literacy as a combination of interaction and awareness of ethical concerns regarding AI. Based on Benjamin Bloom's taxonomy, Ng and colleagues (2021) conceptualized AI literacy as a combination of knowing, comprehension, utilization, assessment, and development. In another paper, Ng et al. (2023) stressed the understanding of principles of fairness, accountability, transparency, ethics, and safety across various applications. Wang et al. (2022) considered the ability to appropriately recognize, utilize, and evaluate AI-based technologies by considering ethical principles as AI literacy. Long and Magerko (2020) also think that the ability to identify ethical issues like fairness and transparency in decisions made by AI-based technologies serves as a marker of individuals' AI literacy. It is not only the ability to operate AI systems but also the ability to use these systems effectively and responsibly, along with the capacity to understand their broader social impacts (Casal-Otero et al., 2023).

Therefore, AI literacy is a combination of abilities to identify applications and limitations, analyze impacts including constraints, recognize the ethical dilemmas involved, and understand the social implications. Besides, literacy is becoming a crucial indicator of determining success in the future. The level of literacy determines the extent to which people would use this technology to solve problems and thereby advance professional success. AI literacy can also play a crucial role in shaping the development trajectory of society through the effective utilization of AI. In journalism specifically, AI literacy is a prerequisite for responsible reporting, preventing harmful impacts on news production (Hollanek et al., 2025). AI literacy enables journalists not only to use these tools responsibly but also to participate in policy debates while maintaining professional and ethical standards.

### Why Bangladesh's Case Matters

While AI literacy is essential globally, its importance is amplified in developing-country contexts like Bangladesh, where resource constraints and limited prior technology adoption create distinct conditions. Journalists can be treated among the early adopters of technologies, the adoption of technology among journalists across different countries is influenced by social, economic, and political factors (Gulyás, 2016; Holman & Perreault, 2023). Similarly, the adoption of technologies, including artificial intelligence, will not be the same for the West and Asian countries like Bangladesh. Difference between technological transmission in Western countries and developing countries is also evident (Wright et al., 2019). Therefore, the use of AI technologies by Bangladeshi journalists might differ from the adoption of AI by journalists in the developed world, such as the West, because of the many differences between Bangladesh and Western countries like the United States of America. For instance, according to World Bank reports, the United States recorded a GDP of $27.72 trillion compared to $437.42 billion in Bangladesh in 2023 (World Bank, 2024a, 2024b). The World Bank reports also demonstrate differences between the two countries in terms of gross domestic product, digital connectivity, fixed broadband use, and mobile cellular subscription rates were also observed. On press freedom index, Bangladesh ranked 149 against U.S. and UK's ranking of 57 and 20 respectively (Reporters Without Borders, 2025). Overall, the socio-economic conditions, technology adoption, and press freedom in Bangladesh are significantly different from those of other developed countries like the U.S. and the U.K., leaving room for journalistic practice and the adoption of technologies like AI to differ—therefore prompting research into AI adoption in such contexts.

Keeping the global AI use and adoption factors in mind, this study focuses on examining Bangladesh's perspective from a foundational level—specifically, what journalists in the country think about AI's potential role in journalism, including its implications and possibilities, and how they are using the technology. These contextual factors—economic constraints, the digital divide, press freedom limitations, and limited prior AI adoption—create conditions fundamentally different from Western newsroom environments. Understanding AI adoption specifically within Bangladesh's journalism context therefore addresses a critical gap in the existing literature, which is largely dominated by studies of well-resourced Western newsrooms.

RQ1: How aware are the journalists in Bangladesh about the scope of Artificial Intelligence use by news media and by journalists in general?

RQ2: How are the journalists in Bangladesh using Artificial Intelligence in their daily activities?

RQ3: How do journalists perceive they can benefit from adopting emerging AI tools?

RQ4: Do journalists identify any negative implications or limitations of using artificial intelligence for journalism?

RQ5: How the news media in Bangladesh as institutions taking stance-relevant to adopting AI?

**Method**

*Data Collection*

To achieve the research objective, we executed a semi-structured qualitative research approach based on approval from the relevant Institutional Review Board for the Protection of Human Subjects (IRB #18503). The interview-based research approach fits the requirements of this paper (Lindlof & Taylor, 2019) as it helps capture context-dependent nuances. The interview protocol involves 17 interview questions. The first question (How do you view artificial intelligence in general? What comes to your mind when you think about artificial intelligence, and why?) was designed to know about the participants level of understanding about AI. Each question was open-ended in nature (e.g.; How can artificial intelligence tools be used in your professional or personal purposes?) (see Appendix A). The interviews were conducted in Bangla, but the Appendix A contains the English version of the questions only (both the Bangla and English version of the questions are approved by the IRB).

*Participants and Sampling*

A purposive sampling approach was adopted (Lindlof & Taylor, 2019; Wimmer & Dominick, 2014) that allowed recruitment of journalists from different layers including staff reporter, senior reporter, special correspondents, photojournalist, copy editor, news editors, and editors from three major forms of news outlets—legacy newspapers, online news portals, and televisions. This paper recruited 23 working journalists with their experience ranging between 3 years to 20 years. Of the participants, there were 8 reporters (staff correspondents, senior correspondents, and special correspondents), 6 copy editors (staff copy editors, senior copy editors, and central-desk editors), 6 news gatekeepers (editors, news editors, and deputy editors), and 3 individuals involved in digital content creation, such as a video editor, digital editor, and photojournalist, representing newspapers, online news portals, and television, respectively (see Table 1). One of the researcher's working experiences in Bangladesh helped recruiting participants using his personal connections. Interviews were conducted via zoom during May-September, 2025 and consents were taken orally as approved by IRB. Interviews and coding were conducted simultaneously. Recruitment continued until data saturation was reached with the 21st participant, and two additional interviews (22nd and 23rd) were conducted to confirm that no new codes emerged. On average, each interview length was approximately 35 minutes depending on participant availability and depth of responses.

**Table 1**

*Job Titles of Participants.*

| Role/Designation | Total |
| --- | --- |
| Reporters (Staff Correspondent, Senior Correspondent, Special Correspondent) | 8 |
| Copy editors (Copy editor, Senior copy editor, Central-desk editor) | 6 |
| Gatekeepers (News Editors, Assistant Editor, Editor) | 6 |
| Video editor, digital editor, photojournalist | 3 |
| Total | 23 |

### Data Analysis

Two researchers transcribed the interview data. Both the researchers open-coded the data on NVivo 14. "In open coding, we go through chunks of data on the basis of its coherent meaning—its standing on its own—not by an arbitrary designation of grammar" (Spiggle, 1994, p. 493). Open coding is also called a "shorthand device" and later helps in developing categories. "Codes are the linkages between data and categories... Usually the code comes before category" (Lindlof and Taylor, 2019, p. 317–318). The coding and analysis of the data were guided by the research questions. The open coding was followed by axial coding to categorizing the codes guided the RQs to make sense of the codes. This resulted in the formation of six primary themes. Subsequently, selective coding was conducted to identify the core category that unified the thematic structure. This three-stage coding progression—open, axial, and selective—allowed us to move systematically from descriptive codes to explanatory categories to an integrated theoretical understanding of AI adoption in Bangladeshi journalism. Disagreements-regarding coding and categorization of codes were resolved through discussions between the two researchers. Data triangulation helped to achieve reliability of the findings. Inclusion of participants from different professional roles (reporters, copy editors, editors), different outlet types (legacy newspapers, online news portals, television), and varying levels of experience (ranging from 3 to 20 years in journalism) held achieving data triangulation, strengthening the robustness of our thematic analysis and reducing the risk of perspective-specific biases. Both the researchers are native Bangla speakers and students at U.S. universities, helped catering benefit in both phases–coding in Bangla and translating the findings in English language. Once the data analysis was completed, the relevant interview quotations were then translated into English for using in the finding and discussion sections. Though, such "piecemeal" translation (Birbili, 2000) has limitations of losing of misrepresentations, but both the researcher being bilingual (native Bangla speakers and studentship in U.S. English only setting) allowed reducing the risk (Ercikan, 1998; Halai, 2007).

### Findings

### AI Integration in Bangladeshi Newsrooms: Adoption Patterns and Mechanisms

Journalists in Bangladesh showed growing awareness about artificial intelligence (AI) and its future potential in news media. Interviewees noted becoming aware of AI tools, particularly ChatGPT through non-official sources like colleagues or friends. Participants acknowledged using AI-driven tools including Google Translate, Google Lens, Google Gemini,

DeepSeek, Grammarly, and Adobe Podcast in recent years (Table 2). Their perceived scope of utilization of AI was widespread, ranging beyond conscious and direct utilization of some AI applications. Participants admitted that they unconsciously rely on AI in their daily professional and personal lives. This initial awareness and growing adoption in the context of Bangladeshi journalists provides an opportunity to examine the key constructs of Unified Theory of Acceptance and Use of Technology (UTAUT). UTAUT (Venkatesh et al., 2003) proposes that technology adoption depends on four primary constructs: performance expectancy, effort expectancy, social influence, and facilitating conditions. Along with presenting the findings, this paper evaluates the relevance of UTAUT constructs in Bangladesh's context.

**Table 2**

*Common AI Tools Use and Purposes.*

| AI Tool Mentioned | Primary User Role/Type | Main Application or Benefit Cited |
|---|---|---|
| Chat GPT | Reporters, copy editors, gatekeepers, video editor, digital editor, photojournalist | Writing/Drafting/Scripting/ Compilation/Summary of data/ Information gathering/Research/Translation/Suggesting Headlines |
| Google Gemini | Senior Reporters, Staff Reporters, Reporters, Photojournalists, Video Editor, Digital Editor | Used (less frequently than ChatGPT), Used for spoken English practice, Photo Editing |
| DeepSeek | Editor, Assistant Editor, Special Correspondent | Used for quick information gathering (sometimes preferred over ChatGPT), Same tasks as ChatGPT |
| Grammarly | Reporters, Copy Editors, (English Media) | Checking and correcting English grammar |
| Google Lens | Reporters, Photojournalists | Translating signs/text, Converting press releases (image) to digital text/Word files |
| Adobe Podcast (AI Tool) | Central Desk Editor, Digital Editor | Used for audio editing for content creation |
| Cloud | Reporters, Copy Editors | Used primarily for language translation |

Among the journalists in Bangladesh believed that almost all journalists in Bangladesh use AI tools for professional and personal purposes in some way or other. Only one participant from a national daily reported that the organization had made AI use mandatory and was providing training support to upskill employee expertise. Except for a very few news outlets, institutional direction on whether to adopt AI was highly absent, creating a clear gap in organizational guidance on managing AI biases and inaccuracies. However, journalists recognize the importance of human validation of information, analysis, translations, or any other outputs.

***Where and How Journalists Employ AI: Usage Patterns and Applications***

Integration of GenAI comes with a range of benefits for Bangladeshi journalists like many western and non-western countries. Usage patterns fall within five primary categories: information gathering and research, scriptwriting and content generation, idea generation and brainstorming, editing and language support, and multimedia assistance. These usage patterns suggest that UTAUT's performance expectancy construct—defined as users' beliefs that AI will enhance their job performance—operates strongly in Bangladesh despite lack of institutional support.

**Information gathering and research**. Journalists—reporters, copy editors, and news editors—can access to large volumes of information, including initial background details on various topics and historical data instantly through GenAI. An editor of an online newspaper highlighted AI's ability to quickly provide historical context with references: "Regarding a 50-year-old history, if I give a command to AI tools like ChatGPT or DeepSeek to organize the information with references, they help me with links" (Participant 4). This finding directly supports UTAUT's performance expectancy construct. According to UTAUT (Venkatesh et al., 2003), high performance expectancy is a primary driver of adoption intention. This utility helps journalists in research or exploring topics outside a journalist's regular beat. A television reporter explicitly connected AI's efficiency to reduced human effort: "I am getting many things through a software [AI] in a short time, my effort will also decrease" (Participant 1).

**Scriptwriting, content generation, and brainstorming.** Journalists reported widespread use of AI for drafting news scripts, social media content, and interview questions. Participants mentioned the application of AI in generating different parts of a news script: "I ask it to change a nice headline, to create an intro for me, to make it into bullet points" (Participant 18). The effort expectancy construct from UTAUT and UTAUT 2—perceived ease of use—appears initially high among Bangladeshi journalists, who found ChatGPT "user-friendly and straightforward". Journalist also finds GenAI effective at mimicking particular writing styles, though with occasional errors: "If I give it a text and direct it to write it in the style of Prothom Alo, it does it, while there might be some errors" (Participant 14). Journalists found AI useful as a partner for generating new ideas or finding multiple news angles when facing writer's blocks. A news editor stated AI's utility: "I am immensely benefited [by using GenAI], especially in the context of my idea generation and when I am helpless, finding nothing" (Participant 22).

**Editing and language support**. In Bangladesh, where English is a second language, journalists found AI highly helpful for translation of English content to Bangla and for checking spelling, proofreading grammar, and sentence structure errors. An assistant editor from a newspaper described this process: "In a small paper like ours, the head of the proof section uses AI in the sense that he asks it to suggest appropriate Bengali spellings, or to help cut a headline" (Participant 10). The appeal of AI for language tasks reveals an important dimension of effort expectancy specific to non-Western journalism. For Bangladeshi journalists, whose work often requires managing multiple languages (Bangla and English), GenAI tools reduce the cognitive effort required for translation and editing. This contrasts with Western journalism contexts where English-monolingual staff may experience lower effort expectancy for these tasks. Thus, effort expectancy operates context-specifically, with non-English-speaking journalists experiencing greater perceived benefit.

**Multimedia assistance**. Journalists in Bangladesh also use AI tools for visual content creation and editing. As a reporter from a newspaper explained: "They are generating images; if you provide news, it creates a picture according to it" (Participant 23). Journalists also pointed out using AI tools for audio content editing as well as creating images based on text. Using AI-generated images helped give online webpages a more professional appearance.

### Effects and Implications of AI Integration: Benefits and Concerns

**Perceived Benefits: Efficiency, Capability, and Professional Necessity.** Efficiency and productivity benefits were highly cited by participants as significant reasons behind adoption of GenAI. These efficiency and productivity benefits align with UTAUT's primary motivation, performance expectancy. However, the Bangladesh context contributes to an important dimension, facilitating conditions. Though UTAUT hypothesizes it as necessary for adoption, the adoption reached to a high level despite the absence of the facilitating condition. Therefore, the professional necessity may override facilitating conditions variables in contexts like Bangladesh. A substantial number of participants stressed these reasons. Journalists across newsrooms acknowledged GenAI as a "time saver" highly effective in increasing work speed and efficiency. One journalist explained: "By giving a command to ChatGPT, you can do it in just five minutes which would previously take 25 minutes. In that case, your 20 minutes are saved" (Participant 1). This testimony exemplifies how performance expectancy matters highly in Bangladesh.

GenAI assistance also enhanced quality by improving journalists' capacity to deal with large volumes of information otherwise might have been impossible to process. AI helped journalists draft news scripts by providing initial information, data, or models, making the process faster and more organized. AI could summarize large content in seconds, create headlines, write introductions, and convert text into bullet points. One senior television reporter believed: "If we use it properly, then it is possible to improve its [content] quality" (Participant 11). This observation demonstrates sophisticated understanding of UTAUT's broader logic, suggesting that in developing-country journalism contexts, performance expectancy encompasses not just output speed but output quality. Moreover, participants also stressed on the ease of use as they find using GenAI tools user-friendly.

Beyond practical benefits, journalists recognized AI as essential for remaining competitive in an increasingly digital and globalized media landscape. They understood that adopting AI was crucial to remain competitive and avoid falling behind peers or other news organizations. This recognition of competitive necessity reveals an important social influence mechanism not fully captured by UTAUT's social influence construct. Rather, Bangladesh journalists report industry-level social influence: awareness that competitors, peers internationally, and the broader journalism field are adopting AI creates implicit pressure to adopt independently. A senior reporter of an online news media emphasized: "if I cannot keep pace [in terms of GenAI adoption], then I will fall behind…Because we need it to survive" (Participant 3).

### Constraints and Concerns: Accuracy, Creativity Loss, and Systemic Risks

Along with benefits, journalists also find downsides of GenAI adoption. The concerns, however, does not significantly impede adoption. This pattern suggests that in Bangladesh, concerns operate independently from adoption decision-making—a finding that differ from Western research (Wenger et al., 2024). This nonrelational dynamics between perceived risk and adoption intention may reflect the "voluntary-compulsion" dynamic proposed for non-Western journalism contexts. In such contexts professional survival needs prevail over risk perception in adoption decisions.

Accuracy and verification emerged as leading concerns among Bangladeshi media professionals. Across almost all participants, inaccuracy, incomplete, or misleading information were identified as dominant drawbacks. A special correspondent from a newspaper pointed out AI's tendency to provide outdated information due to knowledge base frequency bias:

> "When the AI was trained, the frequency of 'former US President' was much higher and that's why when it gives information about the yesterday's bill, which is called the Big Beautiful Bill, it used Trump's title as former president" (Participant 18).

Journalists' articulation of AI hallucination mechanics demonstrates awareness that contrasts with traditional UTAUT assumptions about user knowledge. It means that users, particularly journalists in developing country, may adopt technology despite full understanding. It creates a distinct way of adoption not addressed in UTAUT framework. Apart from these concerns, over-reliance on AI also raised concerns about cognitive miserliness and creativity loss, leading journalists' brains to become "lazy" or "dull," to accomplish tasks like original and critical thinking, in-depth analysis, and creative problem-solving. A newspaper journalist raised this concern directly:

> "The creative thoughts we had before are not happening because the brain is becoming lazy. ChatGPT can give me information based on its information base, but when I think of it as a journalist, I can come up with creative thoughts that AI cannot give me" (Participant 2).

This concern—that reliance on AI diminishes cognitive capability—reflects a value-rationality concern not captured by UTAUT's instrumentally-rational framework. This tension may be particularly acute in resource-constrained contexts where cost pressures push toward instrumental rationality. Relying on AI from early career stages could hinder journalists' ability to develop unique writing styles, investigative skills, and overall journalistic maturity.

The human touch element was viewed as likely to be absent when news content generation became heavily dependent on generative AI. A television channel staff reporter raised this concern: "People can see it and say that it is writing like a machine, and a human element seems to be missing from it" (Participant 1). Moreover, AI lacked understanding of complex societal dynamics, pressure groups, or ethical dilemmas of when to reveal certain truths based on prevailing circumstances, which human journalists navigate. Participants noted AI struggles with Bangla language nuances and cultural context necessary for compelling storytelling. A copy editor underscored: "As a Bangla speaker, it seems to me that the Bangla language undergoes many changes. But it [AI] cannot grasp these changes" (Participant 12). The concern about AI's inability to capture linguistic and cultural nuance represents a critical observation about technology context-fit. UTAUT assumes technology adoption can be analyzed independently of application context, but this finding suggests that in non-Western journalism, language-specific and culture-specific limitations create persistent concerns that don't prevent adoption but create ongoing friction. Fixing this friction involves training that can help journalist to navigate the friction better.

GenAI-based automation capabilities could lead to substantial reduction in human staff, making roles such as proofreaders, graphic designers, translators, and certain reporter types redundant. Which, in turn, can create societal impact of increased unemployment in media. A newsroom editor articulated implications:

> "However, for a very small country like ours, where unemployment is a major problem, if machines keep doing all the work so accurately, there is a high possibility that countless people will become unemployed unless the state brings this into a framework" (Participant 12).

Media professionals raised concerns over AI's capability to generate fake news, propaganda, and deepfakes, posing severe risk of misleading the public and eroding trust in media. There was significant risk of "misuse" or "abuse" of AI for malicious purposes, including creating fake news, rumors, or manipulated content like deepfakes to defame or politically influence individuals. Privacy and data security concerns also emerged prominently. Several participants underscored copyright and plagiarism tensions, noting GenAI could provide information without proper source citations, violating intellectual property rights.

### Institutional Context and Organizational Stance

News media institutions in Bangladesh adopted cautious stance in terms of adopting artificial intelligence either formally or informally. However, none of the participants discussed formal AI policy adoption by news organizations despite widespread organizational knowledge of AI adoption at all levels.

The most significant deviation from UTAUT's facilitating conditions construct in Bangladesh is lack of institutional stance. Facilitating conditions—defined as "organizational and technological infrastructure support system to make use of technology easy and smooth" (Venkatesh et al., 2003)—are largely absent in Bangladesh. None of the participants reports formal institutional AI policies, training programs (except Prothom Alo), and organizational guidelines for managing AI risks (accuracy, bias, plagiarism, misinformation). Yet AI adoption proceeded at high rates. This pattern directly challenges UTAUT's prediction that facilitating conditions significantly predict adoption, suggesting that facilitating conditions may function as a weaker or contingent predictor in developing-country journalism contexts because of professional survival concerns. Furthermore, institutional support does not drive adoption (adoption was already widespread) but rather manages adoption, potentially mitigating risks and improving practice. This suggests a temporal distinction: facilitating conditions may matter less for adoption initiation but more for adoption quality and ethical implementation in later stages.

### Discussion and Conclusion

This study explores the adoption of GenAI in Bangladeshi newsrooms from a foundational level. It examines journalists' awareness, acceptance, usage patterns, and organizational stances toward GenAI. Some of its findings confirms existing findings and some brings new issues to the discourse revolving around the adoption of this technology in the newsrooms. The significance of the findings includes how non-Western, developing contexts differ from Western counterparts and by examining the applicability of the Unified Theory of Acceptance and Use of Technology (UTAUT) in such contexts.

### Similarities in West vs non-West

Apart from these, this study also confirms several previous study findings including the benefits of GenAI—efficiency, productivity, enhanced research capabilities, creative support—along with the concerns over accuracy, cognitive miserliness, job displacement, misinformation risks, privacy vulnerabilities, and societal harm (Newman, 2023; Simon, 2022; Wenger et al., 2024). The concerns were intense in the context of a resource constrained context.

### GenAI Adoption in Non-Western Context: A Different Trajectory

Based on the news media journalists interviewed, one major striking finding was that the news media, except for the leading national daily Prothom Alo, in the Bangladesh does not have any specific policy stance on adoption of AI. However, the use of AI in Bangladesh is widespread. Almost all journalists in Bangladesh use AI for different purposes. Whereas, many Western news media has taken policy stance on using AI tools and techniques. The New York Times, The Washington Post, and Reuters have institutionalized AI integration as the adopted specialized tools, and formal training programs in this respect (Chan-Olmsted, 2019; Jamil, 2020). Therefore, the journalists in Bangladesh engages with AI is motivated by their own individual and professional development motivation. Except for the lone media, AI training is widely absent in the newsrooms in the country. Journalist train themselves on GenAI tools through informal channels like news media, colleagues, friends, and self-learning rather than organizational motivation. Of the participants, participant from a single news media reported to have mandatory AI training for their employees. The disparity in institutional stance is also consistent with the country's socioeconomic disparities with the developed West.

The second major finding is relevant to the high adoption of GenAI despite the socioeconomic disparity. While adoption of the previous technologies involved high cost, this new form of technology does not require come with the cost barrier, leading to high adoption despite the absence of organizational motivation or stance (Helberger & Diakopoulos, 2023). Therefore, journalists were able to adopt AI based on their own utilities.

The absence of organizational AI policies in news media left Bangladeshi journalist to set their own policies when it comes to the question of bias, inaccuracy, and plagiarism concerns. This process of navigation is highly informed by their own ethical principles and rule of thumb–verification and cross-checking. Greater absence of training makes during the adoption process in another key difference between the Western newsrooms where AI literacy training has become increasingly formalized (Beckett, 2019; Wenger et al., 2024) against absence in Bangladesh's news media.

### UTAUT in Non-Western Contexts

One of the key theoretical contributions of this study lies in examining how the Unified Theory of Acceptance and Use of Technology (UTAUT) apply—or fails to apply—in a non-Western, developing country context. The findings reveal significant departures from UTAUT's predictions, suggesting the need for a modified framework for similar contexts. Consistency with The UTAUT theory includes *performance expectancy and effort expectancy, but social influence* works in a horizontal order through informal peer pressure or professional motivation in absence of vertical formal institutional hierarchical or peer pressure. Another critical variable–the facilitating condition–was highly absent in Bangladesh because of the absence of institutional AI policy and guidelines, and formal training, challenging the UTAUT prediction regarding the role of facilitating condition on adoption, particular in voluntary use contexts (Venkatesh et al., 2003). Voluntariness in the context of Bangladeshi journalists is underpinned by their professional compulsion. Meaning, some journalist felt compelled to get use to with GenAI for their own survival which is described as a "competitive necessity". This is consistent with reason behind dropping voluntariness in UTAUT2 (Venkatesh et al., 2012), but this sense of self willingness can come from competitive necessity.

### Proposed Changes for Journalists' Adoption in Developing Countries

Based on these findings, this study proposes partial modification suggestions to the UTAUT framework for non-Western, developing country contexts, based on validation.

1. Facilitating Condition: Absence or weak institutional stance on adoption of news technologies, the role of *facilitating condition* might be a weaker predictor than the UTAUT suggests.
2. Effort Expectancy: The role of *effort expectancy* remains significant throughout the adoption process rather than diminishing over time.
3. Informal Social Influence: Horizontal social influence or peer pressure may play significant role in determining adoption.
4. Voluntary-Compulsion Spectrum: In absence of organizational stance or pressure, a sense of voluntary-compulsion may determine adoption where professional survival pressure is high.

5. Resource Constraint as Moderator: Behavioral intent and use behavior can be moderated by economic constraints as scarcity of resources my hamper exploring advanced features despite high intent to use.
6. Institutional Vacuum as Contextual Factor: Exploration of institutional policy absence can have tested as another moderating factor.

**Limitations and Future Research**

Despite the attempt to contribute the greater discourse on AI adoption in journalism in the context of Bangladesh and to the UTAUT, this paper has limitations. Qualitative research with low sample as a research approach itself lacks generalizability. Therefore, the proposed suggestions to the UTAUT may lack statistical significance if quantitative method applied. However, future studies may consider the proposal to empirically validate its applicability in non-Western context and generalizability.

To move forward as ethic and rule-bound media organizations, Bangladesh and similar non-Western developing countries, should consider developing institutional policy and framework on GenAI. GenAI comes with both benefits and risk which requires careful development and implementation of policies, adequate training, organizational support which altogether enable media professionals to use AI effectively, ethically, and to check biases and inaccuracies.

**Disclosure Statement**

One of the authors previously worked as a journalist in Bangladesh but has no current professional affiliation with any news organization in Bangladesh or elsewhere. This professional experience and native proficiency in Bangla provided valuable contextual understanding but did not influence the study's design, data analysis, or interpretation. The author declares no financial or other competing interests.

## References

Adjin-Tettey, T. D., Muringa, T., Danso, S., & Zondi, S. (2024). The role of artificial intelligence in contemporary journalism practice in two African countries. *Journalism and Media, 5*(3), 846-860. https://doi.org/10.3390/journalmedia5030054

Albizu-Rivas, I., Parratt-Fernández, S., & Mera-Fernández, M. (2024). Artificial Intelligence in Slow Journalism: Journalists' Uses, Perceptions, and Attitudes. *Journalism and Media*, *5*(4), 1836-1850. https://doi.org/10.3390/journalmedia5040111

Beckett, C. (2019). *New powers, new responsibilities: A global survey of Journalism and Artificial Intelligence.* London: The London School of Economics and Political Science, London, UK.

Birbili M. (2000). Translating from one language to another. *Social Research Update, 31* (1–7). https://sru.soc.surrey.ac.uk/SRU31

Broussard, M., Diakopoulos, N., Guzman, A. L., Abebe, R., Dupagne, M., & Chuan, C.-H. (2019). Artificial Intelligence and Journalism. *Journalism & Mass Communication Quarterly*, *96*(3), 673-695. https://doi.org/10.1177/1077699019859901

Brown, T., Mann, B., Ryder, N., Subbiah, M., Kaplan, J. D., Dhariwal, P., ... & Amodei, D. (2020). Language models are few-shot learners. *Advances in neural information processing systems*, *33*, 1877-1901. https://doi.org/10.48550/arXiv.2005.14165

Casal-Otero, L., Catala, A., Fernández-Morante, C., Taboada, M., Cebreiro, B., & Barro, S. (2023). AI literacy in K-12: A systematic literature review. *International Journal of STEM Education, 10*(1), 29. https://doi.org/10.1186/s40594-023-00418-7.

Chan-Olmsted, S. M. (2019). A review of artificial intelligence adoptions in the media industry. *International journal on media management*, *21*(3-4), 193-215. https://doi.org/10.1080/14241277.2019.1695619

Cohen, S., Hamilton, J. T., & Turner, F. (2011). Computational journalism. Communications of the ACM, 54(10), 66-71. https://doi.org/10.1145/2001269.2001288

Cools, H., & Diakopoulos, N. (2024). Uses of Generative AI in the Newsroom: Mapping Journalists' Perceptions of Perils and Possibilities. *Journalism Practice*, 1–19. https://doi.org/10.1080/17512786.2024.2394558

Diakopoulos, N. (2019). *Automating the news: How algorithms are rewriting the media.* Harvard University Press.

Ercikan, K. (1998). Translation effects in international assessments. *International Journal of Educational Research, 29*(6), 543-553. https://doi.org/10.1016/S0883-0355(98)00047-0

Gondwe, G. (2024). Artificial Intelligence, Journalism, and the Ubuntu Robot in Sub-Saharan Africa: Towards a Normative Framework. *Digital Journalism, 13*(4), 826–844. https://doi.org/10.1080/21670811.2024.2311258

GPT-3. (2020, September 8). *A robot wrote this entire article. Are you scared yet, human?* The Guardian. https://www.theguardian.com/commentisfree/2020/sep/08/robot-wrote-this-article-gpt-3

Gulyás, A. (2016). Hybridity and social media adoption by journalists: An international comparison. *Digital Journalism, 5*(7), 884-902. https://doi.org/10.1080/21670811.2016.1232170

Halai N, (2007). Making use of bilingual interview data: Some expressions from the field. *Qualitative Research Report, 12*(3) 344–355. https://ecommons.aku.edu/pakistan_ied_pdck/22

Harb, Z., & Arafat, R. (2024). The Adoption of Artificial Intelligence Technologies in Arab Newsrooms: Potentials and Challenges. *Emerging Media, 2*(3), 371-381. https://doi.org/10.1177/27523543241291068

Helberger, N., & Diakopoulos, N. (2023). ChatGPT and the AI Act. *Internet Policy Review, 12* (4),1-26. https://doi.org/10.14763/2023.1.1682

Hollanek, T., Peters, D., Drage, E., & Hernandes, R. (2025). AI, journalism, and critical AI literacy: exploring journalists' perspectives on AI and responsible reporting. *AI & Society* (2025). https://doi.org/10.1007/s00146-025-02407-6.

Holman, L., & Perreault, G. P. (2023). Diffusion of Innovations in Digital Journalism: Technology, Roles, and Gender in Modern Newsrooms. *Journalism, 24* (5)*.* https://doi.org/10.1177/14648849211073441

Horne, B. D., Nevo, D., O'Donovan, J., Cho, J. H., & Adalı, S. (2019, July). Rating reliability and bias in news articles: Does AI assistance help everyone?. In *Proceedings of the international AAAI conference on web and social media* (Vol. 13, pp. 247-256). https://doi.org/10.1609/icwsm.v13i01.3226

Jamil, S. (2020). Artificial Intelligence and Journalistic Practice: The Crossroads of Obstacles and Opportunities for the Pakistani Journalists. *Journalism Practice*, *15*(10), 1400–1422. https://doi.org/10.1080/17512786.2020.1788412

Lewis, S. C., & Simon, F. M. (2023). Why human-machine communication matters for the study of journalism and artificial intelligence. In A. L. Guzman, R. McEwen, & S. Jones (Eds.), *The SAGE Handbook of Human-Machine Communication* (1st ed.). Sage Publishing.

Lewis, S. C., Zamith, R., & Bunquin, J. B. A. (2025). Technological Hype and AI in Journalism: Five Functions and Why They Matter. *Digital Journalism*, 1–12. https://doi.org/10.1080/21670811.2025.2557994

Lindlof & Taylor (2019). Producing data II: Qualitative interviewing. In *Qualitative Communication Research Methods* (4th ed.). (Chap. 7: pp. 219-275). Los Angeles: Sage.

Long, D., & Magerko, B. (2020). What is AI literacy? Competencies and design considerations. *CHI '20: Proceedings of the 2020 CHI Conference on Human Factors in Computing Systems*, 1–16. https://doi:10.1145/3313831.3376727

Longoni, C., Fradkin, A., Cian, L., & Pennycook, G. (2022, June). News from generative artificial intelligence is believed less. *In Proceedings of the 2022 ACM Conference on Fairness, Accountability, and Transparency* (pp. 97-106). https://doi.org/10.1145/3531146.3533077

Mansoor, H. M., Bawazir, A., Alsabri, M. A., Alharbi, A., & Okela, A. H. (2024). Artificial intelligence literacy among university students—a comparative transnational survey. *Frontiers in Communication, 9*, 1478476. https://doi.org/10.3389/fcomm.2024.1478476

Marikyan, D., & Papagiannidis, S. (2023). Unified theory of acceptance and use of technology: A review. In S. Papagiannidis (Ed.), *TheoryHub book*. Newcastle University. https://open.ncl.ac.uk / ISBN: 9781739604400

McCarthy, J. (2007). From here to human-level AI. *Artificial Intelligence, 171*(18), 1174-1182. https://doi.org/10.1016/j.artint.2007.10.009

McCarthy, J., Minsky, M. L., Rochester, N., & Shannon, C. E. (1955, August 31). A proposal for the Dartmouth summer research project on artificial intelligence. http://raysolomonoff.com/dartmouth/boxa/dart564props.pdf

Newman, N. (2018). *Journalism, media and technology trends and predictions 2018*. Reuters Institute for the Study of Journalism.

Newman, N. (2018). *Journalism, media and technology trends and predictions 2018.* Reuters Institute for the Study of Journalism.

Ng, D. T. K., Lee, M., Tan, R. J. Y., Hu, X., Downie, J. S., & Chu, S. K. W. (2023). A review of AI teaching and learning from 2000 to 2020. Education and Information Technologies, 28(7), 8445–8501. https://doi.org/10.1007/s10639-022-11491-w.

Ng, D. T. K., Leung, J. K. L., Chu, S. K. W., & Qiao, M. S. (2021). Conceptualizing AI literacy: An exploratory review. *Computers and Education: Artificial Intelligence*, 2. 100041. https://doi:10.1016/j.caeai.2021.100041

Nilsson, N. J. (1998). *Artificial intelligence: a new synthesis.* Morgan Kaufmann. Morgan Kaufmann Publishers, Inc.

Noain-Sánchez, A. (2022). Addressing the impact of artificial intelligence on journalism: The perception of experts, journalists, and academics. *Communication & Society, 35*(3), 105–121. https://doi.org/10.15581/003.35.3.105-121

Ogbebor, B., & Carter, C. (2021). Introduction: Innovations, Transitions and Transformations in Digital Journalism: Algorithmic, Symbolic, and New Forms of Journalism. *Digital Journalism, 9*(6), 687–693. https://doi.org/10.1080/21670811.2021.1964854

Onebunne, A. P. (2025). Journalists' Perception of Artificial Intelligence: The Case of Nigerian Newsrooms. *Digital Journalism*, 1–20. https://doi.org/10.1080/21670811.2025.2577127

Ooi, K. B., Tan, G. W. H., Al-Emran, M., Al-Sharafi, M. A., Capatina, A., Chakraborty, A., ... & Wong, L. W. (2025). The potential of generative artificial intelligence across disciplines: Perspectives and future directions. *Journal of Computer Information Systems*, *65*(1), 76-107. https://doi.org/10.1080/08874417.2023.2261010

OpenAI. (2022). Introducing ChatGPT. Retrieved from https://openai.com/blog/chatgpt/

OpenAI. (2025). ChatGPT (Nov 2023 version) [Large language model]. https://chat.openai.com/

Owsley, C. S., & Greenwood, K. (2022). Awareness and perception of artificial intelligence operationalized integration in news media industry and society. *AI & Society, 39*(2), 417–431. https://doi.org/10.1007/s00146-022-01386-2

Rayyan, O. M. (2025). AI and the evolution of journalistic practices. *QScience Jist, 2025*(2). https://doi.org/10.5339/jist.2025.15

Reporters Without Borders. (2025, May 2). World Press Freedom Index 2025: Over half the world's population in "very serious" situation. https://rsf.org/en/world-press-freedom-index-2025-over-half-worlds-population-red-zones

Sengar, S. S., Hasan, A. B., Kumar, S., & Carroll, J. (2024). Generative artificial intelligence: a systematic review and applications. *Multimedia Tools and Applications* 84 (23661-233700 (2025). https://doi.org/10.1007/s11042-024-20016-1

Simon, F. M. (2022). Uneasy Bedfellows: AI in the News, Platform Companies and the Issue of Journalistic Autonomy. *Digital Journalism, 10*(10), 1832–1854. https://doi.org/10.1080/21670811.2022.2063150

Simon, F. M. (2024). *Artificial intelligence in the news: How AI retools, rationalizes, and reshapes journalism and the public arena.* Tow Center for Digital Journalism, Columbia University. https://doi.org/10.7916/ncm5-3v06

Sonni, A. F. (2025). Digital transformation in journalism: mini review on the impact of AI on journalistic practices. *Frontiers in Communication, 10*, 1535156. https://doi.org/10.3389/fcomm.2025.1535156

Spiggle, S. (1994). Analysis and interpretation of qualitative data in consumer research. *Journal of Consumer Research, 21*(3), 491-503. https://doi.org/10.1086/209413

Steinbauer, G., Kandlhofer, M., Chklovski, T., Heintz, F., & Koenig, S. (2021). A differentiated discussion about AI education K-12. *Künstliche Intelligenz, 35*(2), 131–137. https://doi:10.1007/s13218-021-00724-8

Sun, M., Hu, W., & Wu, Y. (2022). Public perceptions and attitudes towards the application of artificial intelligence in journalism: From a China-based survey. *Journalism Practice, 16*(3), 548–570. https://doi.org/10.1080/17512786.2022.2055621

Thomas, R. J., & Thomson, T. J. (2023). What Does a Journalist Look like? Visualizing Journalistic Roles through AI. *Digital Journalism, 13*(4), 631–653. https://doi.org/10.1080/21670811.2023.2229883

Venkatesh, V., & Zhang, X. (2010). Unified Theory of Acceptance and Use of Technology: U.S. Vs. China. *Journal of Global Information Technology Management*, *13*(1), 5–27. https://doi.org/10.1080/1097198X.2010.10856507

Venkatesh, V., Morris, M. G., Davis, G. B., & Davis, F. D. (2003). User Acceptance of Information Technology: Toward a Unified View. *MIS Quarterly*, *27*(3), 425–478. https://doi.org/10.2307/30036540

Venkatesh, V., Thong, J., & Xu, X. (2012). Consumer acceptance and use of information technology: Extending the Unified Theory of Acceptance and Use of Technology. *MIS Quarterly*, *36*(1), 157. https://doi.org/10.2307/41410412

Verma, D. (2024). Impact of artificial intelligence on journalism: A comprehensive review of AI in journalism. *Journal of Communication and Management, 3*(02), 150-156. https://doi.org/10.58966/JCM20243212

Wenger, D., Hossain, M. S., & Senseman, J. R. (2024). AI and the Impact on Journalism Education. *Journalism & Mass Communication Educator 80*(1), https://doi.org/10.1177/107769582412964

Wimmer, R., & Dominick, J. (2014). *Mass Media Research: An Introduction* (10th edition). Belmont, CA: Wadsworth Publishing Co.

World Bank. (2024a). *Bangladesh: Data*. The World Bank. https://data.worldbank.org/country/bangladesh

World Bank. (2024b). *United States: Data*. The World Bank. https://data.worldbank.org/country/united-states

Wu, S. (2024). Journalists as individual users of artificial intelligence: Examining journalists'" value-motivated use" of ChatGPT and other AI tools within and without the newsroom. *Journalism*. https://doi.org/10.1177/14648849241303047.

Yoko, K., (2016, January 13). *What is Needed for AI Literacy?* Research Institute of Economy, Trade and Industry (RIETI). https://www.rieti.go.jp/en/columns/s16_0014.html

Zheng, Y., Zhong, B., & Yang, F. (2018). When algorithms meet journalism: The user perception to automated news in a cross-cultural context. *Computers in human behavior, 86*, 266-275. https://doi.org/10.1016/j.chb.2018.04.046

Funding details:
This work was supported by the *Owen Kulemeka Memorial Award for Research*, Gaylord College of Journalism and Mass Communication, University of Oklahoma.

Generative AI Use Disclosure:
The authors acknowledge the use of generative AI, Claude.ai (Version: Haiku 4.5) and ChatGPT (Version: CPT-5), for language improvement, including checking grammatical errors, organizational structure, and flow.

**Appendix A**
**Semi-Structured Interview Protocol**

Research Question 1*:* How aware are the journalists in Bangladesh about the scope of Artificial Intelligence use by news media and by journalists in general?

Interview Question 1: How do you view artificial intelligence in general? What comes to your mind when you think about artificial intelligence, and why?

Interview Question 2: How long have you known about artificial intelligence tools, and how have you come to know about them (e.g., newspaper, friend, colleagues etc)? And what was your initial thought about them when you first came to know about the AI tools?

Interview Question 3: Have you ever used AI tools like ChatGPT, Gemini, DeepSeek, or any other artificial intelligence tools? If yes, what do you know about them?

Research Question 2: How are the journalists in Bangladesh using Artificial Intelligence in their daily activities?

Interview Question 4: How can artificial intelligence tools be used in your professional purposes (e.g., reporting/copy-editing/photography/editorial decision-making/research/knowledge improvement)?

Interview Question 5: How have you so far used artificial intelligence tools, if at all? If yes, what purposes did it serve for you?

Interview Question 6: Have you heard about any of your friends or colleagues using them for their professional and personal purposes? If yes, how and why do they use them?

Interview Question 7: How easy or difficult have the artificial intelligence tools been so far for you in terms of using them for your professional or personal purposes?

Research Question 3: How do journalists perceive they can benefit from adopting emerging AI tools?

Interview Question 8: How can the use of artificial intelligence tools benefit the news media in general?

Interview Question 9: Do you think it can benefit the quality of journalism in general (e.g., reporting, copy-editing, editorial decision-making)? If yes, how?

Interview Question 10: How can using the AI tools benefit you in the future?

Interview Question 11: How has the use of AI tools been beneficial for your professional causes?

Interview Question 12: How might AI be beneficial for news consumers as well as society in general?

Research Question 4: Do journalists identify any negative implications or limitations of using artificial intelligence for journalism?

Interview Question 13: Do you see any negative consequences AI might bring for journalism in general? If yes, what are they and how?

Interview Question 14: How might it create disadvantages for news reporters/copy editors/news editors/editors? And how?

Interview Question 15: How might it create disadvantages for news consumers as well as society in general?

Research Question 5: How the news media in Bangladesh as institutions taking stance-relevant to adopting AI?

Interview Question 16: What is your organization's institutional stance on adopting AI? In what ways can AI be used to accomplish tasks related to research, reporting, and editing?

Interview Question 17: Does your organization have an AI policy, whether published or unpublished? If yes, could you please provide an overview? If not, why has an AI policy not been formulated so far?